\documentclass{article}

\PassOptionsToPackage{numbers, compress}{natbib}
 \usepackage[preprint]{neurips_2026}

\usepackage[utf8]{inputenc} 
\usepackage[T1]{fontenc}   
\usepackage{hyperref}     
\usepackage{url}           
\usepackage{booktabs}      
\usepackage{amsfonts}       
\usepackage{nicefrac} 
\usepackage{microtype}     
\usepackage{xcolor}         
\usepackage{latexsym}

\usepackage{multirow}
\usepackage{pifont}
\usepackage{amsmath}
\usepackage{amssymb}
\newcommand{\cmark}{\ding{51}}
\newcommand{\xmark}{\ding{55}}
\usepackage{graphicx}
\usepackage{subcaption}
\usepackage{longtable}
\title{PerfCodeBench: Benchmarking LLMs for System-Level High-Performance Code Optimization}

\author {
    {\bf Huihao Jing}\textsuperscript{ \hspace{-0.2em}\includegraphics[height=1em]{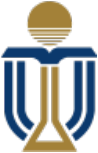}},
    {\bf Wenbin Hu}\textsuperscript{ \hspace{-0.2em}\includegraphics[height=1em]{assets/HKUST.pdf}},
    {\bf Shaojin Chen}\textsuperscript{ \hspace{-0.2em}\includegraphics[height=1em]{assets/HKUST.pdf}},
    {\bf Haochen Shi}\textsuperscript{ \hspace{-0.2em}\includegraphics[height=1em]{assets/HKUST.pdf}},
    {\bf Hanyu Yang}\textsuperscript{
      \hspace{-0.2em}
      \includegraphics[height=1em]{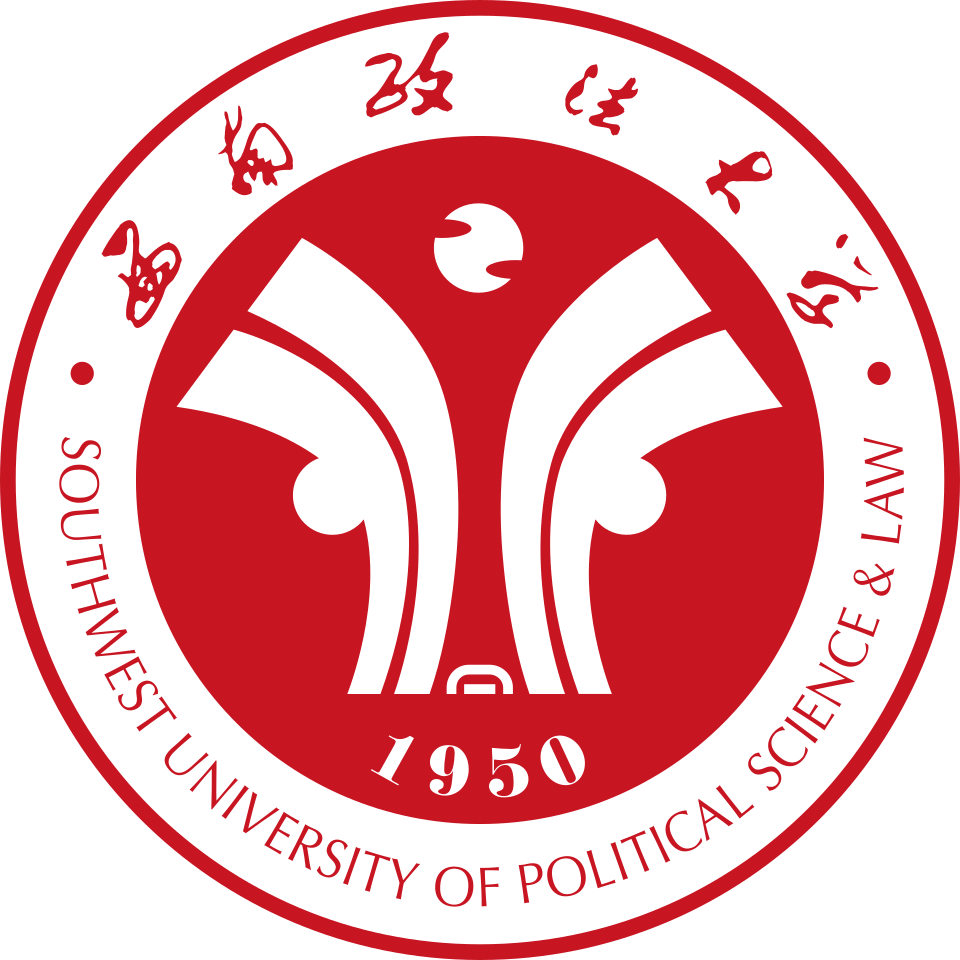}
      \hspace{0.1em}
      \includegraphics[height=1em]{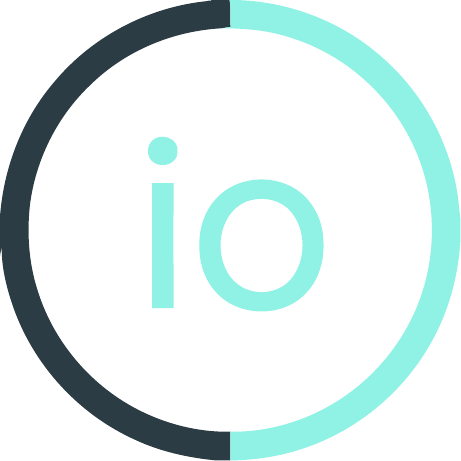}
    },\\
    {\bf Sirui Zhang}\textsuperscript{
      \hspace{-0.2em}
      \includegraphics[height=1em]{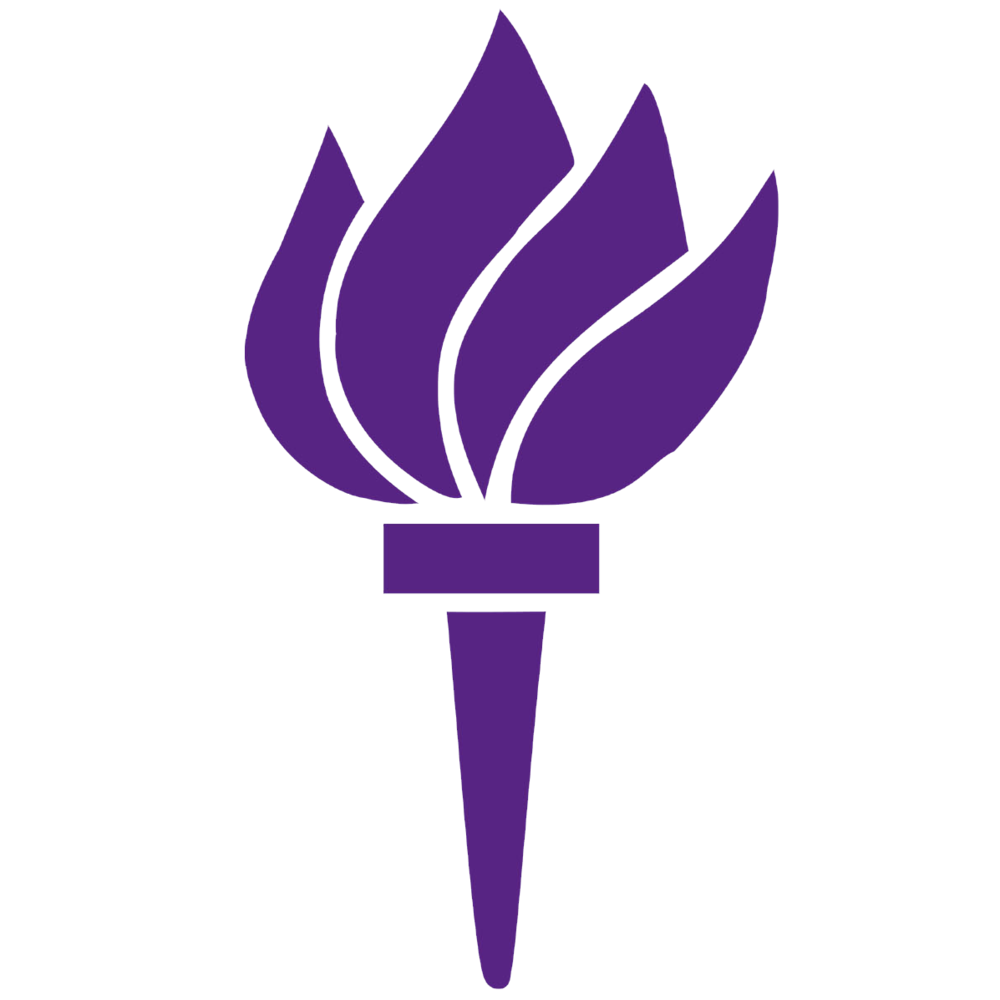}
      \hspace{0.1em}
      \includegraphics[height=1em]{assets/modeio.pdf}
    },
    {\bf Haoran Li}\textsuperscript{
      \hspace{-0.2em}
      \includegraphics[height=1em]{assets/HKUST.pdf}
    }\thanks{Corresponding author}\hspace{0.15em},
    {\bf Yangqiu Song}\textsuperscript{ \hspace{-0.2em}\includegraphics[height=1em]{assets/HKUST.pdf}}\\
    \textsuperscript{\includegraphics[height=1em]{assets/HKUST.pdf}}HKUST, 
    \textsuperscript{\includegraphics[height=1em]{assets/nyu.png}}NYU, 
    \textsuperscript{\includegraphics[height=1em]{assets/swupl.png}}SWUPL,
    \textsuperscript{\includegraphics[height=1em]{assets/modeio.pdf}}MODEIO.AI,\\
    \texttt{hjingaa@connect.ust.hk}\\
}

\begin{document}

\maketitle

\begin{abstract}
Large language models (LLMs) can often generate functionally correct code, but their ability to produce efficient implementations for performance-critical systems tasks remains limited. Existing code benchmarks mainly emphasize correctness or algorithmic problem solving, while realistic systems-level optimization is still underexplored. To address this gap, we introduce PerfCodeBench, an executable benchmark for evaluating LLMs on high-performance code optimization. The tasks require system-level implementation choices, hardware-aware optimization, and careful handling of performance bottlenecks. Each task includes executable correctness checks, a baseline implementation, and a reference optimized solution. This allows us to evaluate both correctness and runtime-oriented efficiency. Our evaluation on a broad set of state-of-the-art LLMs shows a clear gap between model-generated code and expert-optimized implementations. The gap is especially large on tasks involving parallelism and GPU operations. Current models also show weaknesses in cross-language robustness and in consistently reaching expert-level efficiency. These results suggest that performance-aware evaluation are still needed. LLMs should move beyond generating merely correct code toward producing efficient systems software. We submit the benchmark data, evaluation infrastructure, and complete logs of all LLMs-generated code at \href{https://github.com/HKUST-KnowComp/Perfcodebench}{\textcolor{blue}{\underline{repository}}}.
\end{abstract}

\section{Introduction}
Recent frontier Large language models (LLMs), including Claude Opus 4.5~\cite{anthropic2025claudeopus45}, GPT-5.5~\cite{openai2026gpt55}, and DeepSeek-V4~\cite{deepseek2026v4preview}, further demonstrate the rapid progress of LLMs in coding, tool use, and agentic workflows. Systems such as Claude Code~\cite{anthropic2026claudecode}, Codex agents~\cite{openai2026codex}, and OpenClaw~\cite{openclaw2026overview} show their potential to iteratively edit, test, and refine code in practical development workflows. Alongside this progress, recent benchmarks demonstrate strong potential in code generation, from function-level synthesis to repository-level software engineering. Benchmarks such as SWE-bench~\cite{DBLP:conf/iclr/JimenezYWYPPN24}, LiveCodeBench~\cite{DBLP:conf/iclr/JainHGLYZWSSS25}, and BigCodeBench~\cite{DBLP:conf/iclr/ZhuoVCH0WYZHPB025} show that modern models can generate usable code and solve realistic programming tasks. SWE-bench further shows progress on real-world GitHub issue resolution, while contest-style evaluations highlight growing algorithmic reasoning ability. Recent efficiency-oriented studies such as EffiBench~\cite{DBLP:conf/nips/0005QSCZ24}, EffiBench-X~\cite{DBLP:journals/corr/abs-2505-13004} show that LLM-generated programs still fall significantly behind human expert implementations in efficiency, highlighting high-performance code generation as a crucial yet underexplored direction for the next stage of LLM-based coding systems.

However, existing benchmarks and optimization studies still focus mainly on algorithmic efficiency. Their tasks are often drawn from competitive programming, interview-style problems, or exam-level settings. These tasks are useful for testing algorithm selection and data-structure reasoning, but they reveal little about system-level efficiency in real performance-critical software. In practice, efficient code depends on more than asymptotic complexity. It also requires careful control of cache locality, data movement, parallel execution, synchronization cost, hardware utilization and more. This leaves a critical gap in current LLM evaluation. As coding agents become better at editing repositories, running tests, and refining implementations, they also need to optimize code under realistic execution constraints. Without this ability, LLM agents may generate usable code but remain unreliable for performance engineering.

To address this gap, we introduce \textbf{PerfCodeBench}, a benchmark for evaluating LLMs on system-level code optimization. Each task is executable and built around a deterministic harness. Each benchmark instance includes two fixed implementations: a correct but slower baseline and a human-expert optimized reference. Model-generated code is evaluated by the same harness. PerfCodeBench checks compilation, task-specific correctness, and runtime performance. The benchmark covers a broad set of performance-critical domains, including GPU programming, CPU/cache optimization, parallel computing, data processing, AI inference systems and more.

We construct PerfCodeBench from public repositories, benchmark suites, and systems-oriented artifacts, with the full source list provided in Appendix~\ref{app:data_sources}. Candidate tasks are processed and filtered through the following checks: 1. Whether they can compile and run. 2. Whether they have an automatic correctness test. 3. Whether the reference implementation is truly faster than the baseline. 4. Whether the runtime measurements are stable. After filtering, PerfCodeBench keeps tasks that are realistic but still easy to run and reproduce. We use this benchmark to evaluate a wide range of state-of-the-art LLMs. The results show that current models can often generate correct code, but they do not consistently reach expert-level efficiency. This gap is especially clear for parallel programs and GPU kernels.

Our contributions are as follows:
\begin{itemize}
    \item We introduce \textbf{PerfCodeBench}, a large-scale executable benchmark for testing whether LLMs can optimize systems code without changing its semantics. PerfCodeBench has 1,854 executable tasks from real systems sources. The tasks cover six programming languages and a wide range of system tasks.
    \item We conduct a broad evaluation of state-of-the-art LLMs on PerfCodeBench. The results show that current models can often generate correct code, but they still fall short of human-expert optimized implementations. Even the strongest model reaches or surpasses the expert reference on only 61.6\% of comparable tasks. These results reveal a clear gap between current LLM-generated code and human-expert optimized implementations. They also show the need for further research on performance-aware training and evaluation. Such progress is essential for enabling LLMs to generate code that is not only correct, but also efficient across different programming languages and systems workloads.
\end{itemize}

\section{Related Work}

\subsection{LLM-Based Coding Agents}

Recent work has moved beyond standalone code generation toward LLM-based coding agents that can navigate repositories, invoke tools, execute commands, and iteratively revise code. Early agentic software-engineering systems, such as SWE-agent, AutoCodeRover, OpenHands, and Agentless, showed that repository interaction, tool use, test execution, localization, and repair loops are central to practical coding agents~\cite{DBLP:conf/nips/YangJWLYNP24,DBLP:conf/issta/0002RFR24,DBLP:conf/iclr/0001LSXTZPSLSTL25,DBLP:journals/corr/abs-2407-01489}. More recent studies further characterize coding agents as an emerging software-engineering paradigm with greater autonomy, broader task scope, and tighter integration into development workflows~\cite{dong2025surveycodeagents,tang2026agenticse}. Newer work also studies how to scale agent training data, improve agent context, analyze agent trajectories, and support self-evolving software agents~\cite{yang2025swesmith,gloaguen2026evaluatingagentsmdrepositorylevelcontext,bouzenia2025trajectories,xia2025livesweagent}. Beyond research prototypes, practical systems such as Claude Code, Codex agents, and OpenClaw demonstrate this shift in real development workflows~\cite{anthropic2026claudecode,openai2026codex,openclaw2026overview}. Together, these advances suggest that code agents have achieved remarkable progress in automatically completing practical coding tasks.

\subsection{Evaluation of LLM-Generated Code}
\begin{table*}[t]
\centering
\scriptsize
\setlength{\tabcolsep}{3.0pt}
\renewcommand{\arraystretch}{1}
\resizebox{\textwidth}{!}{
\begin{tabular}{l r cccc ccccccc}
\toprule
\multirow{2}{*}{\textbf{Benchmark}} 
& \multirow{2}{*}{\textbf{\#Tasks}}
& \multicolumn{4}{c}{\textbf{Source}} 
& \multirow{2}{*}{\textbf{Corr.}} 
& \multirow{2}{*}{\textbf{Eff.}} 
& \multirow{2}{*}{\textbf{Sys. Eff.}} 
& \multirow{2}{*}{\textbf{Ref.}} 
& \multirow{2}{*}{\textbf{Bott.}} 
& \multirow{2}{*}{\textbf{Dom.}} 
& \multirow{2}{*}{\textbf{Lang.}} \\
\cmidrule(lr){3-6}
& 
& \textbf{Contest} 
& \textbf{Repo} 
& \textbf{PR} 
& \textbf{Suite} 
& & & & & & & \\
\midrule

LiveCodeBench~\cite{DBLP:conf/iclr/JainHGLYZWSSS25}
& 1055
& \cmark & \xmark & \xmark & \xmark
& \cmark & \xmark & \xmark
& \xmark & \xmark & \xmark & \xmark \\

BigCodeBench~\cite{DBLP:conf/iclr/ZhuoVCH0WYZHPB025}
& 1,140
& \cmark & \xmark & \xmark & \xmark
& \cmark & \xmark & \xmark
& \xmark & \xmark & \xmark & \xmark \\

SWE-bench~\cite{DBLP:conf/iclr/JimenezYWYPPN24}
& 2,294
& \xmark & \cmark & \cmark & \xmark
& \cmark & \xmark & \xmark
& \xmark & \xmark & \xmark & \xmark \\

EffiBench~\cite{DBLP:conf/nips/0005QSCZ24}
& 1,000
& \cmark & \xmark & \xmark & \cmark
& \cmark & \cmark & \xmark
& \cmark & \xmark & \xmark & \xmark \\

EffiBench-X~\cite{DBLP:journals/corr/abs-2505-13004}
& 623
& \cmark & \xmark & \xmark & \cmark
& \cmark & \cmark & \xmark
& \cmark & \xmark & \xmark & \cmark \\

\textbf{PerfCodeBench}
& \textbf{1,854}
& \cmark & \cmark & \cmark & \cmark
& \cmark & \cmark & \cmark
& \cmark & \cmark & \cmark & \cmark \\
\bottomrule
\end{tabular}
}
\caption{Comparison of representative LLM code benchmarks. \#Tasks denotes the approximate number of benchmark instances. Source columns indicate whether tasks come from contests, real repositories, performance pull requests, or benchmark suites. Corr. and Eff. denote correctness and efficiency evaluation. Sys. Eff. indicates system-level efficiency evaluation. Ref. denotes human-expert reference solutions. Bott. indicates explicit real systems bottlenecks. Dom. and Lang. indicate multi-domain and multi-language coverage.}
\label{tab:code_benchmark_comparison}
\vspace{-0.1in}
\end{table*}
Existing benchmarks for LLM-generated code have progressed from isolated correctness tests to broader evaluation settings. Recent work improves contamination-aware programming evaluation through continuously updated contest tasks, as in LiveCodeBench~\cite{DBLP:conf/iclr/JainHGLYZWSSS25}. Other benchmarks increase task realism by introducing complex instructions, diverse library/API usage, and project- or repository-level development scenarios~\cite{DBLP:conf/iclr/ZhuoVCH0WYZHPB025,liu2025projecteval,li2025feabench,realbench2026}. Complementary efforts study web-oriented programming, automatic benchmark generation, and online collective evaluation platforms~\cite{xu2025webbench,chou2025autocodebench,du2025codearena}. Efficiency-oriented benchmarks further begin to measure runtime and memory behavior of generated code~\cite{peng2025coffe,DBLP:conf/nips/0005QSCZ24,DBLP:journals/corr/abs-2505-13004}. However, most existing efficiency evaluations still focus on standalone algorithmic tasks, competitive-programming problems, or exam-style settings. Table~\ref{tab:code_benchmark_comparison} compares representative LLM code benchmarks across multiple dimensions. The comparison shows that existing benchmarks still provide limited coverage of system-level efficiency, especially for realistic performance bottlenecks and expert-reference optimization. PerfCodeBench fills this gap with executable, multi-domain, and multi-language systems optimization tasks.

\section{PerfCodeBench}
\begin{figure*}[!t]
\centering
\includegraphics[
    width=\textwidth,
    trim=15 7.5 15 7.5,
    clip
]{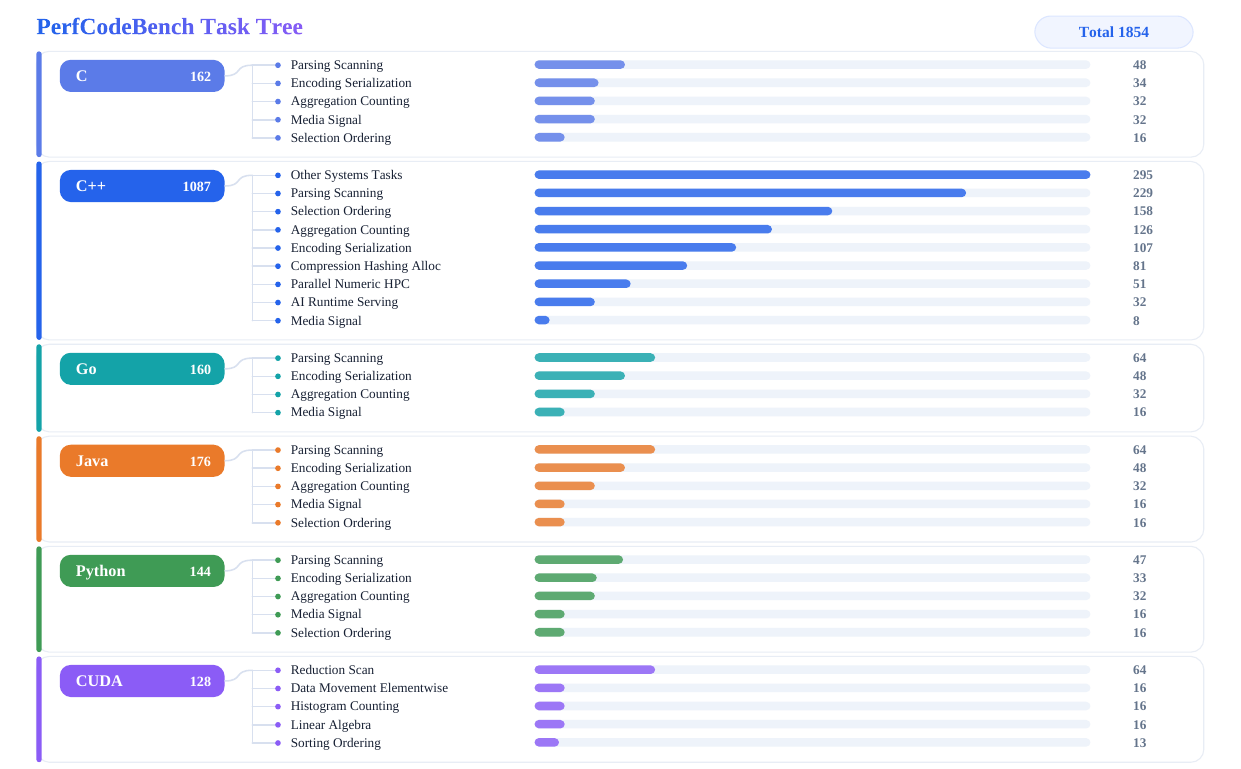}
\caption{Distribution of executable tasks in PerfCodeBench across programming languages and task subdomains.}
\label{fig:task_tree}
\vspace{-0.2in}
\end{figure*}

PerfCodeBench is an executable benchmark for testing whether large language models can optimize systems-oriented code without changing its semantics. Unlike benchmarks that mainly focus on algorithmic coding tasks, PerfCodeBench focuses on performance tasks from realistic software systems. These tasks cover parsing, data processing, compression, SIMD kernels, OpenMP kernels, AI serving components, GPU kernels and more~\cite{DBLP:journals/corr/abs-1902-08318,zstd2026github,lz42026github,xxhash2026github,nvidia2026cudasamples,nvidia2026cub,nvidia2026thrust}. Each task is packaged as a runnable benchmark instance. Each benchmark instance contains a \texttt{baseline} implementation, a human-designed or expert-inspired \texttt{reference} implementation, a correctness oracle, and a runtime-based performance target. PerfCodeBench follows three principles during data collection and task filtering. First, the benchmark should cover diverse tasks. Second, each task should support controlled executable evaluation. Third, each \texttt{reference} implementation should provide a real improvement over the \texttt{baseline}. Benchmark examples can be find in Appendix~\ref{app:bench_examples}.

\subsection{Data Collection}

We build the source pool of PerfCodeBench from public repositories, benchmark suites, and systems research artifacts which is shown in Appendix~\ref{app:data_sources}. At this stage, our goal is to find realistic sources that contain clear performance bottlenecks and can be turned into local, reproducible benchmark tasks. We prioritize sources that are easy to turn into executable tasks. Useful sources usually contain a clear baseline implementation, an optimized implementation or expert patch, an automatic correctness test, or an existing benchmark harness.~\cite{DBLP:journals/corr/abs-1902-08318,yyjson2026github,simdutf2026github,zstd2026github,lz42026github,xxhash2026github,nvidia2026cudasamples}.

The source pool covers several systems domains, matching the categories summarized in Appendix Table~\ref{tab:perfcodebench_data_sources}. These sources include GPU and accelerator benchmarks, CPU SIMD and cache-sensitive libraries, parallel and HPC workloads, SQL and data-processing systems, AI inference runtimes, and general systems libraries. This collection step is intentionally broad. Some sources provide human-written optimized implementations that can be used to construct \texttt{reference} solutions. Others provide benchmark harnesses, representative workloads, or recurring optimization patterns. We do not convert every source directly into a task. Instead, we use these sources as a pool of examples for designing realistic optimization tasks.

\subsection{Task Construction}

After source collection, we convert candidate workloads into executable task instances with a common structure. Each task contains metadata, a benchmark harness, a \texttt{baseline} implementation, a \texttt{reference} implementation, and a writable candidate slot for model-generated code. A key design requirement is interface stability. The \texttt{baseline}, \texttt{reference}, and candidate implementations must follow the same function contract. This allows the harness to test different implementations without changing the benchmark code. Correctness is checked by deterministic task-specific oracles, such as exact checksum equality, exact output equality, exact count equality, or floating-point tolerance thresholds when needed. Performance is measured directly by the harness through repeated runs and median latency. To increase coverage, we expand strong task templates into multiple variants. We vary input size, data density, iteration count, cardinality, or batch shape. The task semantics remain unchanged. This allows PerfCodeBench to scale beyond a small set of manually written tasks while keeping tasks comparable and executable.

\subsection{Task Filtering}

Not every constructed task is kept in the final benchmark. We filter tasks to ensure that they provide meaningful and stable optimization signals.

First, each task must be executable end-to-end. It must compile, run locally, and include an automatic correctness oracle. Second, the \texttt{reference} implementation must be faster than the \texttt{baseline} under the task's own harness. If the \texttt{reference} does not improve over the \texttt{baseline}, the task is removed from the main evaluation pool. Third, we prune highly repetitive task families. Instead of keeping every generated variant, we use family-level caps to preserve representative workload shapes. We also remove or deprioritize tasks with unstable timing, unclear semantics, or brittle packaging. This filtering step turns a large raw task pool (8000+ instances) into a more reliable benchmark with diverse optimization content and controlled executable evaluation (1854 instances).

\section{Scalable Evaluation Pipeline}

Executable evaluation in PerfCodeBench is costly. Each task may require code generation, compilation, correctness checking, and repeated benchmarking. This cost grows quickly when we evaluate thousands of tasks across many models. To make evaluation practical, we use a scalable execution pipeline. It reduces repeated work through caching and schedules tasks based on their workload type. The pipeline combines compile-time caching, benchmark-result caching, and separate CPU/GPU execution pools. This keeps the evaluation faithful while reducing total wall-clock time.

\subsection{Caching and Reuse}

We use caching to avoid repeated computation. First, each benchmark variant is compiled at most once in one evaluation run. The compiled binary is reused if the task source, harness, and implementation do not change. Second, we cache the benchmark results of the \texttt{baseline} and \texttt{reference} implementations. These two implementations are fixed for each task and do not depend on the evaluated model. Therefore, their runtime results only need to be measured once and can be reused across model evaluations. This saves substantial time, especially for C++ and CUDA tasks. Third, we use a two-stage evaluation protocol. We first run a fast screening pass with \texttt{runs=1} on the full task pool. Then, we rerun only correct and promising candidates with more repetitions. This reduces both compilation overhead and benchmark-time cost, making large-scale multi-model evaluation feasible.

\subsection{Split-Pool Scheduling}

We also separate CPU and CUDA tasks during scheduling. A single global parallelism setting is inefficient because CPU and GPU tasks use different resources. CPU tasks usually benefit from high parallelism, since they are often limited by model latency, compilation throughput, and host-side execution. CUDA tasks are different. They share GPU devices, require \texttt{nvcc} compilation, and often have longer benchmarking tails. Running too many CUDA tasks at once can increase contention.

For this reason, we split evaluation into two pools: a high-parallelism CPU pool and a low-parallelism CUDA pool. Each pool has its own worker count. We also use this split during the selective rerun stage. This lets cheap CPU tasks finish quickly while preventing GPU-heavy tasks from slowing down the whole evaluation. This design is useful for PerfCodeBench because its tasks span many domains and have different resource bottlenecks.

\section{Evaluation Metrics}

PerfCodeBench is an executable optimization benchmark. It does not evaluate optimization proposals alone. Each model output must first be executable and correct. Only then do we measure whether it improves performance.

We evaluate each candidate along two dimensions: correctness and efficiency. A candidate must compile, run, and pass the task-specific oracle. If it fails any of these checks, it receives no performance credit. For correct candidates, we compare their runtime with both the \texttt{baseline} and the \texttt{reference}.

\subsection{Correctness and Executability}

For each task instance $i$, let the model-generated implementation be $M_i$. We define correctness as
\begin{equation}
\mathrm{Correct}_i =
\begin{cases}
1, & \text{if } M_i \text{ compiles, runs, and passes the oracle},\\
0, & \text{otherwise.}
\end{cases}
\end{equation}

This is a strict metric. Compilation errors, runtime errors, and wrong outputs are all treated as failures. At the benchmark level, we report the \emph{correct-and-runnable rate}:
\begin{equation}
\mathrm{CRR} = \frac{1}{N} \sum_{i=1}^{N} \mathrm{Correct}_i,
\end{equation}
where $N$ is the number of evaluated tasks. CRR measures how often a model produces valid executable code.

\subsection{Speedup and CGRE}

For each task $i$, let $T^{(i)}_{\text{base}}$, $T^{(i)}_{\text{ref}}$, and $T^{(i)}_{\text{model}}$ be the median runtimes of the \texttt{baseline}, \texttt{reference}, and model-generated implementations. We first define whether the model improves over the \texttt{baseline}:
\begin{equation}
\mathrm{Improve}_i =
\begin{cases}
1, & \text{if } \mathrm{Correct}_i = 1 \text{ and } T^{(i)}_{\text{model}} < T^{(i)}_{\text{base}},\\
0, & \text{otherwise.}
\end{cases}
\end{equation}

We also report raw speedup in the logs:
\begin{equation}
\mathrm{Speedup}_i = \frac{T^{(i)}_{\text{base}}}{T^{(i)}_{\text{model}}},
\end{equation}
which is only used when the candidate is correct.

Raw speedup does not show how close the model is to the expert \texttt{reference}. Therefore, we use \emph{Correctness-Gated Relative Efficiency} (CGRE):
\begin{equation}
\mathrm{CGRE}_i =
\mathrm{Correct}_i \times
\mathrm{clamp}\!\left(
\frac{T^{(i)}_{\text{base}} - T^{(i)}_{\text{model}}}
{T^{(i)}_{\text{base}} - T^{(i)}_{\text{ref}}},
\, 0,\, 1
\right),
\end{equation}
where $\mathrm{clamp}(x,0,1)=\min(1,\max(0,x))$.

CGRE gives zero credit to incorrect outputs and outputs slower than the baseline. It gives full credit when the model matches or beats the \texttt{reference}. Otherwise, it gives partial credit based on how much of the \texttt{baseline}-to-\texttt{reference} gap the model closes.

\subsection{Aggregate Statistics}

We summarize model performance with several benchmark-level metrics. The \emph{faster-than-baseline rate} is
\begin{equation}
\mathrm{FBR} = \frac{1}{N} \sum_{i=1}^{N} \mathrm{Improve}_i.
\end{equation}
FBR measures how often the model produces correct code that is faster than the \texttt{baseline}.

We report the \emph{reference-or-better rate}:
\begin{equation}
\mathrm{RBR} = \frac{1}{N} \sum_{i=1}^{N}
\mathbf{1}\!\left[T^{(i)}_{\text{model}} \leq T^{(i)}_{\text{ref}} \land \mathrm{Correct}_i = 1\right].
\end{equation}
RBR measures how often the model reaches or beats the \texttt{reference}.

Finally, we report thresholded CGRE:
\begin{equation}
\mathrm{CGRE}_{\ge \tau} =
\frac{1}{|\mathcal{C}|} \sum_{i \in \mathcal{C}} \mathbf{1}\!\left[\mathrm{CGRE}_i \ge \tau\right],
\end{equation}
where we use $\tau = 0.8$. This measures how often the model closes at least $80\%$ of the available expert improvement. Together, these metrics capture correctness, speedup frequency, expert-level performance, and performance-gap closure.

\section{Evaluation}
\paragraph{Testbed.}
PerfCodeBench is evaluated in a local executable environment. Each task is compiled and run directly by its benchmark harness. The current benchmark includes 1,854 executable tasks across C, C++, Go, Java, Python, and CUDA.

For each task, the \texttt{baseline}, \texttt{reference}, and LLM-generated candidate are evaluated under the same harness. Performance is measured from the runtime reported by the task. To improve stability and reduce cost, we use adaptive per-task timeouts. Each task is first profiled with the \texttt{baseline} implementation. The candidate timeout is then set using a baseline-dependent multiplier, with fixed lower and upper bounds. This avoids using a single global timeout. Task runtimes vary widely, and a few long-tail tasks can easily slow down the whole parallel evaluation. For more details about testbed, please refer to Appendix~\ref{app:exper_details}.

\paragraph{Models.}
As detailed in Table~\ref{tab:perfcodebench_results}, we evaluate a diverse set of proprietary and open-weight LLMs through an OpenAI-compatible inference interface. The evaluated models include GPT-5.4 and GPT-5~\cite{openrouter2026gpt54,openrouter2025gpt5}; Claude Opus 4.5 and Claude Sonnet 4.5~\cite{anthropic2025claudeopus45,anthropic2026claudemodels}; Gemini 3.1 Pro and Gemini 3.1 Flash~\cite{google2026gemini31pro,google2026geminimodels}; DeepSeek-V4-Pro and DeepSeek-V4-Flash~\cite{deepseek2026v4preview}; Qwen3.6-Max, Qwen3.6-Plus, Qwen3.6-27B, and Qwen3.6-35B-A3B~\cite{qwen2026qwen36,qwen2026qwen3635ba3b}; Gemma-4-31B and Gemma-4-26B-A4B~\cite{google2026gemma4,google2026gemma4overview}; Llama-4-Maverick and Llama-4-Scout~\cite{meta2025llama4,llama2026models}; Kimi K2.6 and Kimi K2~\cite{moonshot2026kimi,moonshot2026k2platform}; and Doubao-Seed-2.0-Lite and Doubao-Seed-2.0-Mini~\cite{bytedance2026seed20,openrouter2026seed20mini}.

\begin{table*}[t]
\centering
\scriptsize
\setlength{\tabcolsep}{4pt}
\resizebox{\textwidth}{!}{
\begin{tabular}{lccccc}
\toprule
\textbf{Model Name} 
& \textbf{CRR(\%)} 
& \textbf{FBR(\%)} 
& \textbf{RBR(\%)} 
& \textbf{CGRE(\%)} 
& \textbf{CGRE$_{\ge 0.8}$(\%)} \\
\midrule
GPT-5.4 & \textbf{71.25} & \textbf{66.72} & \textbf{\textit{51.82}} & 66.27 & \textbf{\textit{64.26}} \\
GPT-5 & 61.81 & 59.55 & \textbf{61.60} & \textbf{73.99} & \textbf{72.91} \\
\midrule
Claude Opus 4.5 & \textbf{\textit{70.55}} & \textbf{\textit{65.75}} & 43.23 & 65.09 & 63.17 \\
Claude Sonnet 4.5 & 69.36 & 62.08 & 32.81 & 61.60 & 59.15 \\
\midrule
Gemini 3.1 Pro & 25.57 & 23.73 & 40.09 & 61.46 & 60.67 \\
Gemini 3.1 Flash & 51.83 & 46.98 & 22.02 & 45.89 & 43.92 \\
\midrule
DeepSeek-V4-Pro & 54.48 & 51.13 & 47.95 & 64.30 & 62.88 \\
DeepSeek-V4-Flash & 47.09 & 44.82 & 43.05 & 61.25 & 60.29 \\
\midrule
Qwen3.6-Max & 47.25 & 42.50 & 37.19 & \textbf{\textit{66.32}} & 63.10 \\
Qwen3.6-Plus & 45.20 & 39.75 & 29.70 & 56.46 & 54.85 \\
Qwen3.6-27B & 4.26 & 3.99 & 39.60 & 65.07 & 61.39 \\
Qwen3.6-35B-A3B & 32.09 & 27.94 & 22.51 & 44.38 & 41.26 \\
\midrule
Gemma-4-31B & 42.45 & 37.65 & 16.91 & 36.69 & 35.48 \\
Gemma-4-26B-A4B & 45.15 & 39.64 & 16.67 & 38.13 & 35.87 \\
\midrule
Llama-4-Maverick & 29.40 & 24.38 & 20.30 & 43.72 & 40.26 \\
Llama-4-Scout & 6.53 & 4.69 & 11.07 & 19.60 & 16.61 \\
\midrule
Kimi K2.6 & 3.94 & 3.29 & 52.13 & 62.14 & 61.70 \\
Kimi K2 & 36.03 & 32.90 & 27.58 & 44.73 & 43.58 \\
\midrule
Doubao-Seed-2.0-Lite & 47.57 & 43.80 & 40.68 & 60.19 & 58.99 \\
Doubao-Seed-2.0-Mini & 1.89 & 1.56 & 18.42 & 34.01 & 34.21 \\
\bottomrule
\end{tabular}
}
\caption{Overall evaluation results on PerfCodeBench. CRR denotes the correct-and-runnable rate. FBR measures the fraction of tasks where the model produces a correct implementation faster than the baseline. RBR measures the fraction of comparable tasks where the model matches or exceeds the reference implementation. CGRE reports correctness-gated relative efficiency, and CGRE$_{\ge 0.8}$ reports the fraction of comparable tasks that close at least 80\% of the baseline-to-reference performance gap.}
\label{tab:perfcodebench_results}
\vspace{-0.2in}
\end{table*}
\subsection{Main Results}

\paragraph{Proprietary LLMs:}
GPT-5.4 and GPT-5 are the strongest proprietary models, but they are strong in different ways. GPT-5.4 gets the best CRR ($71.25\%$) and FBR ($66.72\%$), so it is the most reliable at producing runnable speedups. GPT-5 does better on reference-level efficiency, with the best RBR ($61.60\%$), CGRE ($73.99\%$), and $\mathrm{CGRE}_{\ge 0.8}$ ($72.91\%$). Claude Opus 4.5 is close on CRR and FBR, but weaker on RBR. Claude Sonnet 4.5 is slightly behind Opus. Gemini 3.1 Pro is different again: its CRR is low, but its RBR and CGRE are still decent. This shows that correctness and efficiency are not the same skill.
\paragraph{Open-source LLMs:}
DeepSeek-V4-Pro is the strongest balanced open-source model. It gets $54.48\%$ CRR and $64.30\%$ CGRE. DeepSeek-V4-Flash is close behind. The Qwen family is more mixed. Qwen3.6-Max has a strong CGRE of $66.32\%$, but its CRR is only $47.25\%$. Qwen3.6-27B is even more extreme: it rarely succeeds, but its successful outputs can still be strong. Gemma sits in the middle range. Llama is more polarized, with Maverick somewhat competitive and Scout much weaker. Kimi K2.6 also shows the same pattern of low CRR but decent conditional quality. Doubao-Seed-2.0-Lite is mid-tier, while Doubao-Seed-2.0-Mini is near the bottom.
\paragraph{Overall Patterns:}
Three patterns stand out. First, PerfCodeBench is still far from saturated. The gap between the best and worst models is large on every metric. Second, correctness and efficiency only partly overlap. A model can be good at producing runnable code, but still miss the \texttt{reference} speed. Third, reliability and conditional strength are different. Some models fail often, but their successful outputs can still be strong. The language split shows the same thing. CPU languages are much easier than CUDA. This is why PerfCodeBench needs multiple metrics. CRR alone is not enough, and CGRE alone is not enough either. We report more evaluation results in Appendix~\ref{app:exper_results}.

\subsection{Comparison on Different Language Subsets}

Table~\ref{tab:language_subset_comparison} compares the main language slices in PerfCodeBench. The benchmark is dominated by C++ tasks, which account for $58.63\%$ of the executable pool, while the remaining workload is distributed across C, Go, Java, Python, and CUDA. The results show a clear language-level split. C++ is the easiest large subset, with both high average CRR and high average CGRE across strong models. Go and Java are also relatively tractable, and Claude Opus 4.5 gives the best CGRE on both subsets. Python has the highest average CRR among the strong models, suggesting that executable success is comparatively easier there, but its average CGRE is still below the strongest C++ results. C is noticeably harder than the higher-level CPU subsets. CUDA is the most difficult language slice by a large margin: its average CRR and CGRE are far below all CPU languages. This shows that optimization ability is strongly language-dependent, and that performance-oriented code generation cannot be summarized by a single overall score alone.

\begin{table}[t]
\centering
\small
\setlength{\tabcolsep}{6pt}
\renewcommand{\arraystretch}{1.08}
\begin{tabular}{lcccc}
\toprule
\textbf{Language} & \textbf{Task Share(\%)} & \textbf{Avg. CRR} & \textbf{Avg. CGRE} & \textbf{Best CGRE Model} \\
\midrule
C & 8.74 & 45.16 & 31.33 & GPT-5.4 \\
C++ & 58.63 & 74.73 & 69.20 & GPT-5.4 \\
Go & 8.63 & 76.67 & 71.14 & Claude Opus 4.5 \\
Java & 9.49 & 77.27 & 71.36 & Claude Opus 4.5 \\
Python & 7.77 & 85.53 & 71.87 & DeepSeek-V4-Flash \\
CUDA & 6.74 & 10.53 & 6.97 & DeepSeek-V4-Pro \\
\bottomrule
\end{tabular}
\vspace{0.1in}
\caption{Comparison across language subsets. Average scores are computed over six strong representative models: GPT-5.4, GPT-5, Claude Opus 4.5, Claude Sonnet 4.5, DeepSeek-V4-Pro, and DeepSeek-V4-Flash.}
\label{tab:language_subset_comparison}
\vspace{-0.15in}
\end{table}

\subsection{Performance Gap Across CPU and GPU Tasks}

Table~\ref{tab:cpu_gpu_gap_comparison} shows the gap between CPU and GPU tasks. The pattern is very consistent across models. All representative models perform much better on CPU tasks than on CUDA tasks, both in correctness and in relative efficiency. For example, GPT-5.4 reaches $82.06\%$ average CPU CRR but only $2.40\%$ GPU CRR, and its CPU CGRE is $68.67\%$ while its GPU CGRE is $0.00\%$. Claude Opus 4.5 and GPT-5 are somewhat more robust on GPU than GPT-5.4, but they still remain far below their CPU performance. Among the models listed here, DeepSeek-V4-Pro gives the strongest GPU-side numbers, but even it remains much weaker on GPU than on CPU. This confirms that GPU optimization is a distinct regime in PerfCodeBench. Solving CPU systems code does not automatically transfer to accelerator kernels, where models must handle a different set of bottlenecks such as parallel decomposition, launch structure, synchronization, and memory hierarchy.

\begin{table}[t]
\centering
\small
\setlength{\tabcolsep}{6pt}
\renewcommand{\arraystretch}{1.08}
\begin{tabular}{lcccc}
\toprule
\textbf{Model} & \textbf{CPU CRR} & \textbf{GPU CRR} & \textbf{CPU CGRE} & \textbf{GPU CGRE} \\
\midrule
GPT-5.4 & 82.06 & 2.40 & 68.67 & 0.00 \\
GPT-5 & 71.02 & 9.60 & 64.74 & 6.70 \\
Claude Opus 4.5 & 79.68 & 12.00 & 69.47 & 10.26 \\
Claude Sonnet 4.5 & 75.44 & 9.60 & 60.65 & 5.76 \\
DeepSeek-V4-Pro & 62.04 & 18.40 & 57.18 & 11.99 \\
DeepSeek-V4-Flash & 61.00 & 11.20 & 57.18 & 7.12 \\
\bottomrule
\end{tabular}
\vspace{0.1in}
\caption{Average CPU versus GPU performance for representative strong models. CPU averages are computed over C, C++, Go, Java, and Python; GPU corresponds to CUDA.}
\label{tab:cpu_gpu_gap_comparison}
\end{table}

\subsection{Correctness-Efficiency Decoupling}

\begin{figure}[t]
\centering
\begin{subfigure}{0.49\linewidth}
    \centering
    \includegraphics[width=\linewidth]{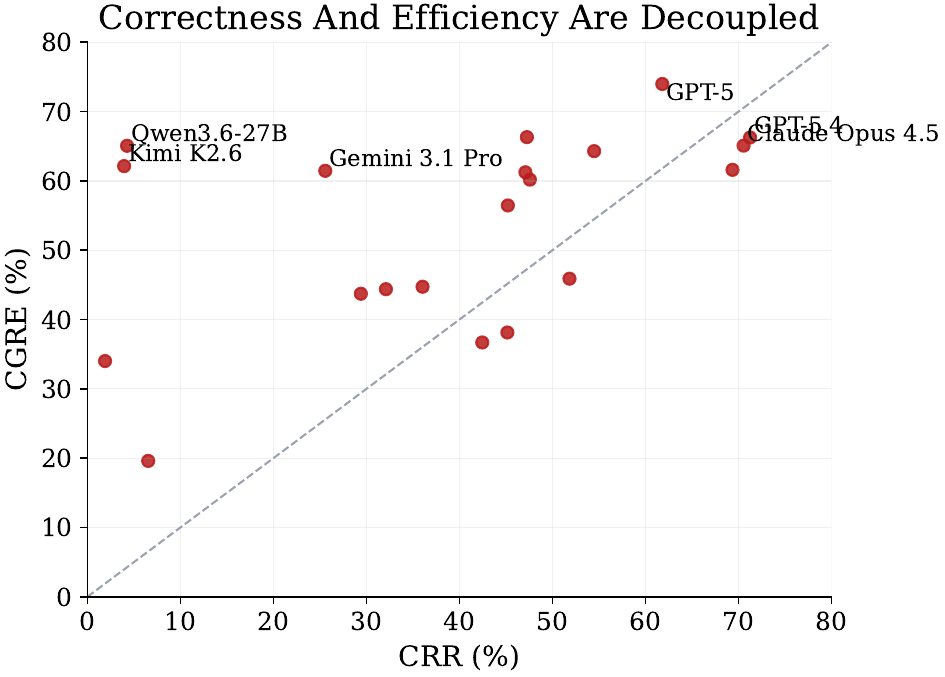}
    \caption{Correctness--efficiency decoupling.}
    \label{fig:finding_decoupling}
\end{subfigure}
\hfill
\begin{subfigure}{0.49\linewidth}
    \centering
    \includegraphics[width=\linewidth]{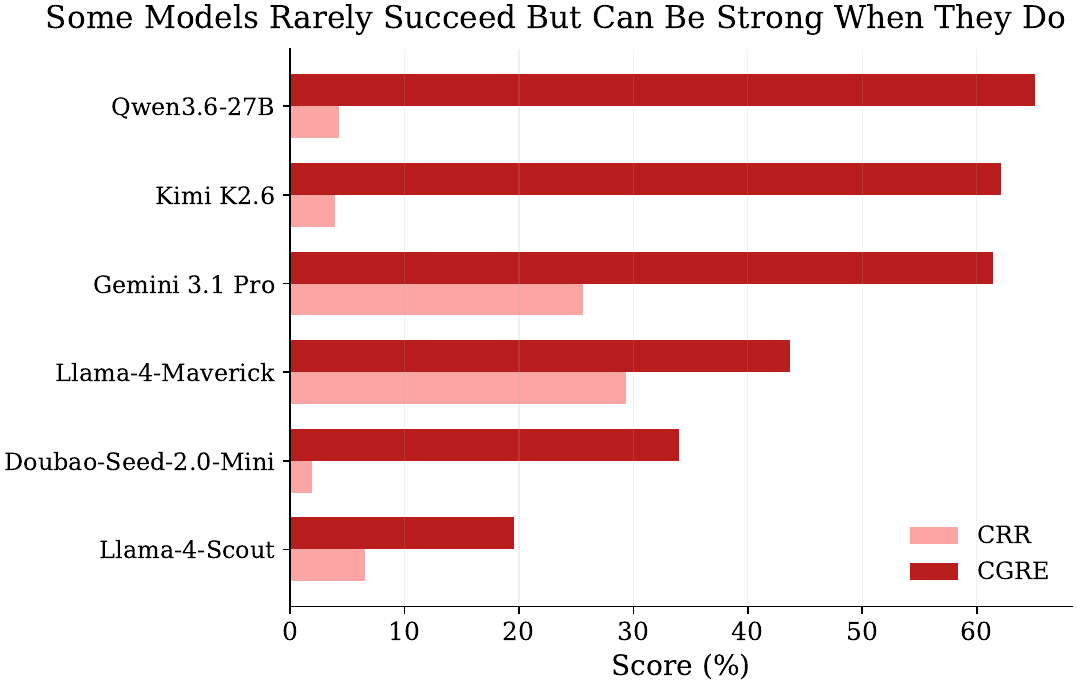}
    \caption{Rare successes.}
    \label{fig:finding_rare_successes}
\end{subfigure}
\caption{Additional findings from PerfCodeBench. Some models rarely succeed, but their successful outputs can still be strong; meanwhile, correctness and efficiency capture different axes of performance.}
\label{fig:additional_findings}
\vspace{-0.25in}
\end{figure}

Figure~\ref{fig:finding_decoupling} shows that correctness and efficiency are related but not identical. GPT-5.4 has the best CRR, but GPT-5 is stronger on RBR and CGRE. This means a model can be better at producing runnable speedups without being the best at closing the expert performance gap. The same pattern appears in other families too. Claude is strong on correctness, but not always on reference-level efficiency. Some open-source models are weaker overall, yet still show strong conditional efficiency when they succeed. This decoupling is one reason we report several metrics instead of a single score.

\subsection{Rare but Strong Successes}

Figure~\ref{fig:finding_rare_successes} shows another pattern: some models fail often, but their successful outputs can still be strong. Gemini 3.1 Pro, Qwen3.6-27B, and Kimi K2.6 all have low CRR, yet their CGRE values are much higher than their success rates might suggest. In these cases, the model does not succeed reliably, but when it does, the result can still be close to the \texttt{reference}. This is useful to keep in mind when judging open-ended optimization models. Low success rate does not always mean low optimization quality. It often means the model is unstable and needs better reliability.

\subsection{Scaling Improves Efficiency More Than Reliability}

A useful pattern is that stronger models within the same family often improve reference-level efficiency more than correctness. This is visible in both the GPT and Claude families. GPT-5.4 has the best CRR, but GPT-5 is stronger on RBR and CGRE. Claude Opus 4.5 and Claude Sonnet 4.5 are also fairly close on CRR, yet the gap becomes larger on the efficiency-oriented metrics. This suggests that model scaling is especially helpful for closing the baseline-to-reference performance gap, not just for making code run.

\section{Conclusion}
PerfCodeBench takes a step toward evaluating whether LLMs can optimize real systems code under executable correctness constraints. Our results suggest that current strong models can often produce runnable and baseline-improving code, but expert-level optimization remains difficult. This is especially true for harder settings such as CUDA and other hardware-sensitive workloads. The benchmark also shows that correctness and efficiency should be studied together. A model may generate valid code without closing much of the performance gap, or occasionally produce a strong optimization while failing often. We therefore believe that executable systems optimization remains an open and challenging problem, and that more realistic evaluation settings are still needed.

\section*{Limitations}
PerfCodeBench has several limitations. First, the benchmark still depends on the quality and coverage of the current task pool. Although we include a broad range of executable workloads, some task families are larger and better represented than others. Second, the benchmark is sensitive to implementation details of the evaluation environment, including hardware, compiler behavior, and runtime setup. Third, the current version focuses on local executable tasks and does not yet cover larger distributed systems, long-horizon optimization workflows, or full application-scale engineering tasks. Finally, some of the current baseline and reference implementations are benchmark-internal constructions rather than directly lifted from a single upstream source, which improves executability but may reduce ecological fidelity in some cases.

\section*{Impact Statement}
PerfCodeBench is intended to support research on practical code optimization rather than surface-level code generation. In this setting, a useful model should preserve correctness while also improving runtime or throughput in measurable ways. We hope the benchmark helps researchers better understand where current models succeed, where they still fail, and which optimization regimes remain most challenging. At the same time, we do not view the current benchmark as a final or complete solution. It should be understood as an evolving evaluation framework that can be expanded with broader workloads, stronger references, and more diverse execution environments in future work.


\newpage
\bibliographystyle{plainnat}
\bibliography{custom}

\clearpage
\appendix

\section{Data Sources}

\label{app:data_sources}

This appendix lists public sources used to build PerfCodeBench. These sources provide realistic systems workloads. They also provide executable benchmark designs and optimization motifs for task construction.

The source pool covers several types of performance-critical code. It includes GPU kernels, CPU SIMD code, cache-sensitive routines, parallel workloads, data-processing systems, AI inference systems, and systems libraries. Many sources contain human-written optimized implementations, which we use to construct the \texttt{reference} solutions in PerfCodeBench. A single source may contribute more than one task family. Other sources provide benchmark harnesses or recurring optimization patterns.

Table~\ref{tab:perfcodebench_data_sources} summarizes the source pool. For each source, we report its domain, implementation language, role in PerfCodeBench, and reference link. The table is organized by source rather than by task instance.

\begin{scriptsize}
\setlength{\tabcolsep}{4pt}
\renewcommand{\arraystretch}{1.05}

\begin{longtable}{p{2.2cm} p{1.8cm} p{1.8cm} p{3.0cm} p{4.0cm}}

\toprule
\textbf{Source} & \textbf{Domain} & \textbf{Languages} & \textbf{Role in PerfCodeBench} & \textbf{Reference Link} \\
\midrule
\endfirsthead

\toprule
\textbf{Source} & \textbf{Domain} & \textbf{Languages} & \textbf{Role in PerfCodeBench} & \textbf{Reference Link} \\
\midrule
\endhead

\midrule
\multicolumn{5}{r}{\emph{Continued on next page}}\\
\endfoot

\bottomrule
\endlastfoot

KernelBench & GPU / Accelerator & Python, CUDA & Public kernel-generation benchmark suite; used as a source of GPU optimization motifs and executable kernel tasks. & \url{https://github.com/ScalingIntelligence/KernelBench} \\
TritonBench & GPU / Accelerator & Python, Triton & Source of operator-level GPU kernels and real-world Triton optimization tasks. & \url{https://github.com/thunlp/TritonBench} \\
ComputeEval & GPU / Accelerator & CUDA, Python & Source of CUDA code-generation and performance-oriented benchmark problems. & \url{https://github.com/NVIDIA/compute-eval} \\
NVIDIA CUTLASS & GPU / Accelerator & C++, CUDA & Source of GEMM, grouped GEMM, and kernel-tuning motifs for accelerator tasks. & \url{https://github.com/NVIDIA/cutlass} \\
NVIDIA CUDA Samples & GPU / Accelerator & C++, CUDA & Source of canonical transpose, reduction, histogram, and matrix multiplication kernels. & \url{https://github.com/NVIDIA/cuda-samples} \\
FlashAttention & GPU / Accelerator / AI Systems & CUDA, C++, Python & Source of memory-bound attention and fused GPU kernel motifs. & \url{https://github.com/Dao-AILab/flash-attention} \\
xFormers & GPU / Accelerator / AI Systems & Python, C++, CUDA & Source of memory-efficient attention and high-performance operator implementations. & \url{https://github.com/facebookresearch/xformers} \\

\midrule
simdjson & CPU SIMD / Cache & C++ & Source of parsing and scanning tasks with explicit SIMD and cache-sensitive optimizations. & \url{https://github.com/simdjson/simdjson} \\
simdutf & CPU SIMD / Cache & C++ & Source of UTF, ASCII, and base64 conversion tasks with vectorized implementations. & \url{https://github.com/simdutf/simdutf} \\
fast\_float & CPU SIMD / Cache & C++ & Source of numeric parsing tasks centered on fast scalar and SIMD conversion paths. & \url{https://github.com/fastfloat/fast_float} \\
yyjson & CPU SIMD / Cache & C & Source of JSON parsing and serialization tasks with low-overhead implementations. & \url{https://ibireme.github.io/yyjson/} \\
Google Highway & CPU SIMD / Cache & C++ & Source of portable SIMD patterns for arithmetic, thresholding, and vectorized reductions. & \url{https://github.com/google/highway} \\
LZ4 & Systems Libraries / SIMD & C & Source of compression and throughput-oriented systems tasks. & \url{https://github.com/lz4/lz4} \\
Zstandard & Systems Libraries / SIMD & C & Source of compression and decompression benchmark motifs with strong performance tuning. & \url{https://github.com/facebook/zstd} \\

\midrule
Rodinia & Parallel CPU / HPC & C, C++, CUDA, OpenMP, OpenCL & Source of shared-memory and accelerator parallel kernels such as hotspot and pathfinder. & \url{https://github.com/yuhc/gpu-rodinia} \\
NAS Parallel Benchmarks & Parallel CPU / HPC & C, Fortran & Source of classic HPC kernels such as CG and MG. & \url{https://www.nas.nasa.gov/software/npb.html} \\
PolyBench/C & Parallel CPU / HPC & C & Source of stencil and dense numerical kernels such as Jacobi-2D and Heat-3D. & \url{https://github.com/ferrandi/PolyBenchC} \\
PARSEC & Parallel CPU / HPC & C, C++, Fortran & Source of shared-memory application motifs and multicore performance workloads. & \url{https://github.com/cirosantilli/parsec-benchmark} \\
SPLASH-2 & Parallel CPU / HPC & C & Source of classic parallel application kernels for shared-memory systems. & \url{https://github.com/staceyson/splash2} \\
Graph500 & Parallel CPU / HPC & C & Source of irregular graph-processing and memory-access-heavy workloads. & \url{https://github.com/graph500/graph500} \\

\midrule
DuckDB & SQL / Data Processing & C++, SQL & Major source of relational query, join, aggregation, and pushdown optimization motifs. & \url{https://github.com/duckdb/duckdb} \\
DuckDB TPC-H / TPC-DS Extensions & SQL / Data Processing & SQL, C++ & Source of analytical SQL tasks and standardized query templates. & \url{https://duckdb.org/docs/stable/core_extensions/tpch} \\
Polars & SQL / Data Processing & Rust, Python & Source of dataframe-style columnar processing and vectorized data transformation tasks. & \url{https://github.com/pola-rs/polars} \\
Apache DataFusion & SQL / Data Processing & Rust & Source of query execution, expression evaluation, and benchmark-backed analytical workloads. & \url{https://github.com/apache/datafusion} \\
ClickHouse & SQL / Data Processing & C++ & Source of high-throughput analytics engine optimizations and query-processing motifs. & \url{https://github.com/ClickHouse/ClickHouse} \\
Velox & SQL / Data Processing & C++ & Source of vectorized execution and data-processing runtime tasks. & \url{https://github.com/facebookincubator/velox} \\
pandas & SQL / Data Processing & Python, Cython, C & Source of row-wise versus vectorized dataframe optimization patterns. & \url{https://github.com/pandas-dev/pandas} \\

\midrule
vLLM & AI Inference Systems & Python, C++, CUDA, Triton & Source of KV-cache, batching, prefix caching, and inference-runtime optimization motifs. & \url{https://github.com/vllm-project/vllm} \\
SGLang & AI Inference Systems & Python, CUDA, Triton & Source of serving-time scheduling and speculative decoding optimization patterns. & \url{https://github.com/sgl-project/sglang} \\
llama.cpp & AI Inference Systems & C, C++ & Source of lightweight inference runtime and cache-management optimization tasks. & \url{https://github.com/ggerganov/llama.cpp} \\
ggml & AI Inference Systems & C, C++ & Source of low-level tensor runtime and backend kernel optimization motifs. & \url{https://github.com/ggml-org/ggml} \\
TensorRT-LLM & AI Inference Systems & C++, Python, CUDA & Source of inference-stack kernel fusion and runtime performance tasks. & \url{https://github.com/NVIDIA/TensorRT-LLM} \\
PyTorch & AI Inference Systems & Python, C++, CUDA & Source of custom operator and extension-based performance-critical workloads. & \url{https://github.com/pytorch/pytorch} \\
ONNX Runtime & AI Inference Systems & C++, C, Python & Source of transformer inference and runtime-level optimization tasks. & \url{https://github.com/microsoft/onnxruntime} \\

\midrule
RocksDB & Systems Libraries & C++ & Source of storage-engine, API, and I/O performance optimization motifs. & \url{https://github.com/facebook/rocksdb} \\
jemalloc & Systems Libraries & C & Source of allocator and memory-management optimization patterns. & \url{https://github.com/jemalloc/jemalloc} \\
mimalloc & Systems Libraries & C & Source of allocator, sharding, and low-overhead memory management motifs. & \url{https://github.com/microsoft/mimalloc} \\
Redis & Systems Libraries & C & Source of low-latency systems programming and pipelining-related motifs. & \url{https://github.com/redis/redis} \\
nginx & Systems Libraries & C & Source of networking and request-processing performance motifs. & \url{https://github.com/nginx/nginx} \\
Protobuf & Systems Libraries & C++, C, Python, Java & Source of serialization and deserialization performance tasks. & \url{https://github.com/protocolbuffers/protobuf} \\
Cap'n Proto & Systems Libraries & C++ & Source of zero-copy serialization and systems-level encoding tasks. & \url{https://github.com/capnproto/capnproto} \\

\midrule
SWE-Perf & Repo-level Performance Benchmark & Python & Source of repository-level performance-improvement task design ideas and evaluation framing. & \url{https://github.com/SWE-Perf/SWE-Perf} \\
SWE-fficiency & Repo-level Performance Benchmark & Python & Source of repository-level performance debugging and optimization task framing. & \url{https://swefficiency.com/} \\

\end{longtable}

\vspace{-0.8em}
\captionof{table}{Main data sources used to construct the PerfCodeBench task pool. The table is organized at the source level rather than the individual task level. The reference link points to the original repository, benchmark suite, or project page from which task motifs, benchmark designs, or executable optimization patterns were derived.}
\label{tab:perfcodebench_data_sources}

\end{scriptsize}

\section{Benchmark Examples}
\label{app:bench_examples}

PerfCodeBench tasks are constructed around concrete code transformations rather than abstract optimization descriptions. Each instance contains a runnable \texttt{baseline}, a stronger \texttt{reference}, and a model-produced \texttt{candidate} that is evaluated under the same harness. This appendix gives several representative examples by showing the actual code patterns that appear in the benchmark. For brevity, each example highlights the key optimization-relevant fragment rather than the full source file.

\subsection{Sorted As-Of Join}

This task asks the model to speed up a sorted as-of join while preserving an exact weighted sum. The main optimization opportunity is to exploit monotonicity in the probe stream instead of repeatedly performing binary search.

\paragraph{Baseline.}
\begin{verbatim}
for (std::size_t i = 0; i < probe_ts.size(); ++i) {
  const auto it = 
        std::upper_bound(dim_ts.begin(), dim_ts.end(), probe_ts[i]);
  const std::size_t idx = 
        static_cast<std::size_t>(it - dim_ts.begin() - 1);
  sum += static_cast<uint64_t>(dim_values[idx]) *
         static_cast<uint64_t>(probe_weights[i]);
}
\end{verbatim}

\paragraph{Reference.}
\begin{verbatim}
std::size_t j = 0;
for (std::size_t i = 0; i < probe_ts.size(); ++i) {
  while (j + 1 < dim_ts.size() && dim_ts[j + 1] <= probe_ts[i]) {
    ++j;
  }
  sum += static_cast<uint64_t>(dim_values[j]) *
         static_cast<uint64_t>(probe_weights[i]);
}
\end{verbatim}

\paragraph{Candidate (GPT-5.4).}
\begin{verbatim}
const uint32_t* dim_ts_data = dim_ts.data();
const uint32_t* dim_val_data = dim_values.data();
const uint32_t* probe_ts_data = probe_ts.data();
const uint32_t* probe_w_data = probe_weights.data();

std::size_t j = 0;
for (std::size_t i = 0; i < probe_n; ++i) {
  const uint32_t pts = probe_ts_data[i];
  while (j + 1 < dim_n && dim_ts_data[j + 1] <= pts) {
    ++j;
  }
  local_sum += static_cast<uint64_t>(dim_val_data[j]) *
               static_cast<uint64_t>(probe_w_data[i]);
}
\end{verbatim}

\begin{table}[t]
\centering
\small
\setlength{\tabcolsep}{6pt}
\renewcommand{\arraystretch}{1.05}
\begin{tabular}{lc}
\toprule
\textbf{Variant} & \textbf{Median Time (ns)} \\
\midrule
Baseline & 976193492 \\
Reference & 88828599 \\
Candidate (GPT-5.4) & 61549189 \\
\bottomrule
\end{tabular}
\caption{Measured performance for \texttt{asof\_join\_weighted\_sum}.}
\label{tab:appendix_asof_perf}
\end{table}

\subsection{Python Top-\textit{k} Selection}

This Python task is representative of higher-level language cases in the benchmark. The baseline fully sorts all scored items, whereas the optimized versions reduce work by keeping only the top-\textit{k} set.

\begin{table}[t]
\centering
\small
\setlength{\tabcolsep}{6pt}
\renewcommand{\arraystretch}{1.05}
\begin{tabular}{lc}
\toprule
\textbf{Variant} & \textbf{Median Time (ns)} \\
\midrule
Baseline & 11760471748 \\
Reference & 1074475213 \\
Candidate (GPT-5.4) & 29007162 \\
\bottomrule
\end{tabular}
\caption{Measured performance for \texttt{python\_topk\_selection\_checksum\_v3603\_000}.}
\label{tab:appendix_python_topk_perf}
\end{table}

\paragraph{Baseline.}
\begin{verbatim}
pairs = sorted(((s, i) for i, s in enumerate(scores)),
               key=lambda x: (-x[0], x[1]))
for score, index in pairs[:k]:
  h ^= ((score << 32) ^ index)
\end{verbatim}

\paragraph{Reference.}
\begin{verbatim}
heap = []
for index, score in enumerate(scores):
  item = (score, -index)
  if len(heap) < k:
    heapq.heappush(heap, item)
  elif item > heap[0]:
    heapq.heapreplace(heap, item)
out = sorted(((score, -neg_index) for score, neg_index in heap),
             key=lambda x: (-x[0], x[1]))
\end{verbatim}

\paragraph{Candidate (GPT-5.4).}
\begin{verbatim}
heap = []
for i, s in enumerate(sc):
  item = (s, -i)
  if len(heap) < k:
    push(heap, item)
  elif item > heap[0]:
    replace(heap, item)
heap.sort(key=lambda x: (-x[0], -x[1]))
\end{verbatim}

\subsection{Java JSON Field Scanning}

This task captures a common data-processing optimization pattern: replacing a convenient but heavyweight parsing method with a tighter task-specific scan that preserves exact output values.

\begin{table}[t]
\centering
\small
\setlength{\tabcolsep}{6pt}
\renewcommand{\arraystretch}{1.05}
\begin{tabular}{lc}
\toprule
\textbf{Variant} & \textbf{Median Time (ns)} \\
\midrule
Baseline & 145661086 \\
Reference & 24158671 \\
Candidate (GPT-5.4) & error \\
\bottomrule
\end{tabular}
\caption{Measured performance for \texttt{java\_json\_field\_checksum\_v2700\_064}.}
\label{tab:appendix_java_json_perf}
\end{table}

\paragraph{Baseline.}
\begin{verbatim}
Matcher m = P.matcher(row);
m.matches();
int u = Integer.parseInt(m.group(1));
int s = Integer.parseInt(m.group(2));
boolean f = m.group(3).equals("true");
\end{verbatim}

\paragraph{Reference.}
\begin{verbatim}
int up = row.indexOf("\"u\":") + 4;
int sp = row.indexOf("\"s\":") + 4;
int u = parseIntAt(row, up);
int s = parseIntAt(row, sp);
boolean f = row.indexOf("\"f\":true") >= 0;
\end{verbatim}

\paragraph{Candidate (GPT-5.4).}
\begin{verbatim}
int idx = 6; // after {"u":
int u = 0;
while ((c = row.charAt(idx)) != ',') {
  u = u * 10 + (c - '0');
  idx++;
}
idx += 5; // skip ,"s":
\end{verbatim}

\subsection{CUDA Matrix Transpose}

This example shows the GPU side of PerfCodeBench. The baseline performs a direct transpose from global memory, while the optimized versions stage values through shared memory and use a padded tile to avoid bank conflicts.

\begin{table}[t]
\centering
\small
\setlength{\tabcolsep}{6pt}
\renewcommand{\arraystretch}{1.05}
\begin{tabular}{lc}
\toprule
\textbf{Variant} & \textbf{Status} \\
\midrule
Baseline & ok \\
Reference & timeout \\
Candidate (GPT-5.4) & error \\
\bottomrule
\end{tabular}
\caption{Observed evaluation status for \texttt{cuda\_transpose\_tiled\_f32} under the current timeout setting.}
\label{tab:appendix_cuda_transpose_status}
\end{table}

\paragraph{Baseline.}
\begin{verbatim}
__global__ void transpose_naive_kernel(...) {
  const int x = blockIdx.x * kTileDim + threadIdx.x;
  const int y = blockIdx.y * kTileDim + threadIdx.y;
  for (int j = 0; j < kTileDim; j += kBlockRows) {
    if (x < width && y + j < height) {
      output[x * height + (y + j)] = input[(y + j) * width + x];
    }
  }
}
\end{verbatim}

\paragraph{Reference.}
\begin{verbatim}
__shared__ float tile[kTileDim][kTileDim + 1];
tile[threadIdx.y + j][threadIdx.x] = input[(y + j) * width + x];
__syncthreads();
output[(ty + j) * height + tx] = tile[threadIdx.x][threadIdx.y + j];
\end{verbatim}

\paragraph{Candidate (GPT-5.4).}
\begin{verbatim}
__shared__ float tile[kTileDim][kTileDim + 1];
#pragma unroll
for (int j = 0; j < kTileDim; j += kBlockRows) {
  const int y = y_in + j;
  if (x_in < width && y < height) {
    tile[threadIdx.y + j][threadIdx.x] = input[y * width + x_in];
  }
}
\end{verbatim}

These examples show the intended use of PerfCodeBench. The benchmark does not only ask whether a model can write correct code. Instead, it tests whether the model can identify the real bottleneck in an executable implementation and rewrite the code in a way that measurably improves runtime under a fixed correctness oracle. The same three-part structure appears throughout the benchmark, but the concrete transformations vary widely across CPU kernels, parsing routines, high-level language implementations, and GPU code.

\section{More Experiment Details}
\label{app:exper_details}

\subsection{More Testbed Details}

PerfCodeBench is evaluated on a local multi-GPU server with two \texttt{Intel Xeon Platinum 8480C} CPUs, 224 logical CPU cores, and 2.0~TiB of RAM. The machine contains three \texttt{NVIDIA H800} GPUs, each with 80~GiB of memory. The software stack used in our current environment includes Python 3.13.3, GCC/G++ 11.4.0, Go 1.18.1, Java 11.0.30, CUDA 12.1, PyTorch 2.8.0, NumPy 2.2.6, Pandas 3.0.2, Numba 0.61.2, and the OpenAI Python client 2.32.0. PerfCodeBench itself runs each task variant through the same build-and-measure protocol, and the median elapsed time is used as the primary runtime statistic. To keep evaluation stable, timeout limits are calibrated per task from baseline profiling rather than fixed globally. This is important because the benchmark includes both lightweight parsing routines and much heavier OpenMP or GPU kernels.

\subsection{Cost Analysis}

The main evaluation cost comes from repeated code execution rather than from the benchmark definition itself. Cost therefore scales with three factors: the number of tasks, the number of repeated runs per task, and the latency of the underlying runtime or compiler toolchain. Tasks that are CPU-bound and short-running are comparatively cheap, while GPU kernels, large Python workloads, and tasks with expensive compilation or warm-up phases are more costly. To limit wasted compute, PerfCodeBench uses adaptive timeouts derived from baseline timing, so slow tasks are truncated early and short tasks are not over-allocated time. In practice, this makes the benchmark much more scalable than a single conservative global timeout would be.

For API-based candidate generation, the total cost is also moderate in our current setup. The code-generation prompt for each task contains only task metadata, the interface contract, and the baseline implementation, and the model returns a single replacement source file in one shot. Across the current 1,854-task benchmark, the total prompt text is about 2.94 million characters and the collected model outputs are about 2.50 million characters. Using a conservative character-to-token approximation and the current standard GPT-5.4 API pricing (\$2.50 per million input tokens and \$15.00 per million output tokens), a full single-model generation sweep remains comfortably below a \$50 budget, and in practice is typically much closer to tens of dollars than to hundreds. Even after accounting for retries, malformed responses, and partial reruns, the current pipeline is still inexpensive enough to support repeated large-scale experiments without making API spend the dominant bottleneck.

\subsection{Prompt}

For the final executable evaluation, the model is prompted to directly generate a replacement implementation for the task rather than to produce a high-level optimization proposal. Table~\ref{tab:perfcodebench_prompt_template} shows the prompt template used during candidate generation. The system message enforces a strict JSON schema, while the user message provides the task goal, metric, correctness rule, optional interface contract, and the full baseline implementation. This design keeps the instruction format simple, while ensuring that the model sees the exact code it must optimize under a fixed task interface.

\begin{table*}[t!]
\small
\centering
\begin{tabular}{p{0.96\textwidth}}
\toprule
\textbf{PerfCodeBench Executable Evaluation Prompt Template} \\
\textbf{System message:} \\
You are a performance engineer. Return JSON only as an object with exactly two string fields: \texttt{"summary"} and \texttt{"solution\_source"}. \\
\midrule
\textbf{User message template} \\
Optimize the following implementation for performance while preserving correctness. Return a full, complete, compilable replacement for \texttt{\textcolor{black}{<solution\_file>}}. Do not change the externally required function signature or entrypoint used by the harness. You may use only the dependencies already compiled by this task. \\
\textbf{Task metadata and code context:} \\
\texttt{Task ID:} \texttt{\textcolor{black}{<task\_id>}} \\
\texttt{Title:} \texttt{\textcolor{black}{<title>}} \\
\texttt{Goal:} \texttt{\textcolor{black}{<goal>}} \\
\texttt{Metric:} \texttt{\textcolor{black}{<metric>}} \\
\texttt{Correctness rule:} \texttt{\textcolor{black}{<correctness\_rule>}} \\
\texttt{Allowed external includes:} \texttt{\textcolor{black}{<include\_list>}} \\
\texttt{Interface / task contract:} \texttt{\textcolor{black}{<interface\_text>}} \\
\texttt{Current baseline <solution\_file>:} \texttt{\textcolor{black}{<baseline\_source>}} \\
\midrule
\textbf{Required output format} \\
\{ \\
\hspace*{1em} \texttt{"summary"}: \texttt{"\textcolor{black}{<brief optimization summary>}"}, \\
\hspace*{1em} \texttt{"solution\_source"}: \texttt{"\textcolor{black}{<full replacement source file>}"} \\
\} \\
\bottomrule
\end{tabular}
\caption{Prompt template used in the final executable evaluation of PerfCodeBench. The model is asked to directly synthesize a replacement source file for the baseline implementation, and the returned code is then compiled and benchmarked under the task harness.}
\label{tab:perfcodebench_prompt_template}
\end{table*}

\section{More Evaluation Results}
\label{app:exper_results}

\begin{figure*}[!t]
\centering
\includegraphics[
    width=\textwidth,
]{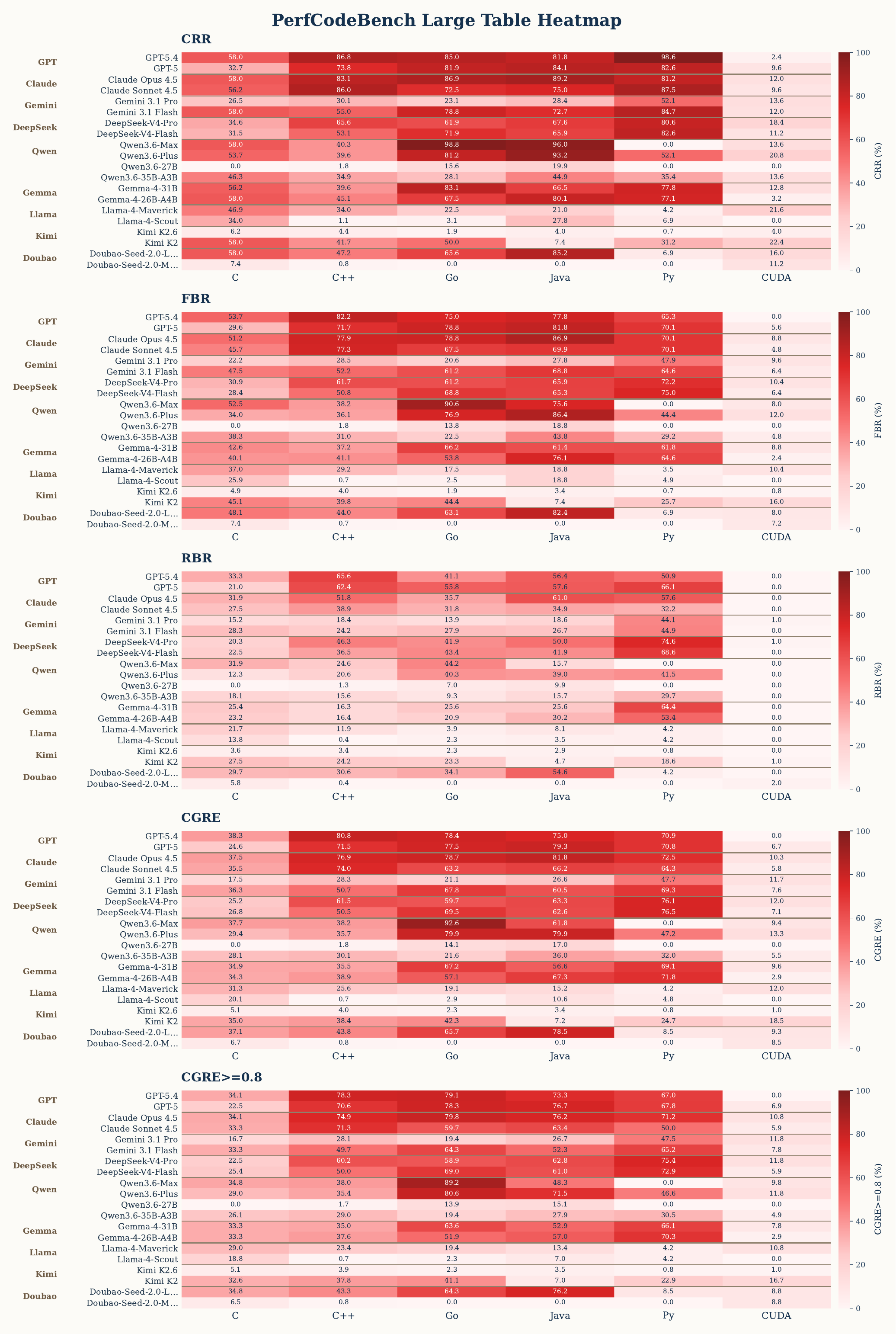}
\caption{Heatmap of the per-language PerfCodeBench results. Rows correspond to evaluated models, grouped by family, and columns correspond to C, C++, Go, Java, Python, and CUDA. Separate panels report CRR, FBR, RBR, CGRE, and CGRE$_{\ge 0.8}$, with darker colors indicating stronger performance.}
\label{fig:large_heat_maps}
\vspace{-0.2in}
\end{figure*}

\subsection{Reading the Heatmap}

Figure~\ref{fig:large_heat_maps} provides a more fine-grained view of model behavior across the major language subsets in PerfCodeBench. Unlike the main table, which reports only benchmark-level aggregates, the heatmap exposes how each model behaves on C, C++, Go, Java, Python, and CUDA separately. This view is useful because overall averages can hide strong domain asymmetries. A model that appears competitive in aggregate may still perform very unevenly across languages, and a model that looks weak overall may still have narrow strengths on specific subsets.

\subsection{Cross-Language Variation}

The most immediate pattern in the heatmap is that model performance is highly language-dependent. C++ remains the strongest large subset for most frontier models, and Go, Java, and Python are also comparatively tractable for the better-performing systems. By contrast, C is more mixed, with stronger sensitivity to model family and optimization style. These differences are consistent with the construction of the benchmark: higher-level CPU subsets often expose clear local optimization opportunities, while lower-level systems tasks more often require tighter handling of memory access, parsing structure, or low-level loop behavior.

\subsection{CPU Versus CUDA}

The heatmap also makes the CPU--GPU gap especially clear. Across nearly all model families, the CUDA columns are much weaker than the CPU-language columns, not only for CRR but also for FBR, RBR, and CGRE. In other words, the difficulty is not limited to getting GPU code to compile and run; even correct CUDA outputs often fail to recover much of the reference-level speedup. This reinforces the conclusion that GPU optimization should be treated as a distinct regime rather than as a straightforward extension of CPU systems optimization.

\subsection{Family-Level Differences}

A second pattern is that model families differ not only in absolute quality but also in how balanced their behavior is across languages. GPT and Claude models show the most uniformly strong CPU-side performance, whereas several open-weight families exhibit a more uneven profile. Some families produce isolated strong results on selected languages while remaining weak elsewhere. The heatmap is helpful here because it distinguishes broad robustness from narrow specialization: a model with a few dark cells may still be much less reliable than a model whose performance is consistently strong across most language subsets.

\subsection{Metric-Specific Effects}

Finally, the heatmap shows that the ranking of models can change across metrics even within the same language. Some models are dark on CRR and FBR but lighter on RBR or CGRE, indicating that they often generate runnable speedups without consistently approaching \texttt{reference} quality. Others show the opposite pattern: they succeed less often, but their successful outputs can be relatively strong. This metric-level separation further supports the use of multiple evaluation criteria in PerfCodeBench rather than relying on a single scalar score.


\end{document}